%% file: main.tex
\documentclass[preprint,10pt]{elsarticle}
\usepackage{graphicx}% Include figure files
\usepackage{dcolumn}% Align table columns on decimal point
\usepackage{amsmath, amsthm, amssymb, amsfonts}
\usepackage{overpic,color}
\usepackage{bm}% bold math
\usepackage{lineno}
\journal{Physics Letters B}
\newcommand{ \Num }[2]{\mathrm{N}\mean{#1}_{#2}}
\newcommand{ \Den }[2]{\mathrm{D}\mean{#1}_{#2}}

\newcommand{ \la }{\langle}
\newcommand{ \ra }{\rangle}

\newcommand{ \mean }[1]{\la #1 \ra}

\newcommand{\bef}{\begin{figure}}
\newcommand{\eef}{\end{figure}}

\newcommand{\be}{\begin{equation}}
\newcommand{\ee}{\end{equation}}
\newcommand{\bea}{\begin{eqnarray}}
\newcommand{\eea}{\end{eqnarray}}

\begin{document}

\begin{frontmatter}

\title{Correlation Measurements Between Flow Harmonics in Au+Au Collisions at RHIC}

\include{author}

%\date{\today}
\begin{abstract}
Flow harmonics ($v_n$) in the Fourier expansion of the azimuthal distribution of particles are widely used to quantify the anisotropy in particle emission in high-energy heavy-ion collisions. The symmetric cumulants, $SC(m,n)$, are used  to measure the correlations between different orders of flow harmonics. These correlations are used to constrain the initial conditions and the transport properties of the medium in theoretical models. In this Letter, we present the first measurements of the four-particle symmetric cumulants in Au+Au collisions at $\sqrt{s_{NN}}$  = 39 and 200 GeV from data collected by the STAR experiment at RHIC.  We observe that $v_{2}$ and  $v_{3}$  are anti-correlated in all centrality intervals with similar correlation strengths from 39 GeV Au+Au to 2.76 TeV Pb+Pb (measured by the ALICE experiment). The $v_{2}$-$v_{4}$ correlation seems to be stronger at 39 GeV than  at higher collision energies. The initial-stage anti-correlations between second and third order eccentricities are sufficient to describe the measured correlations between $v_{2}$ and $v_{3}$. The best description of $v_{2}$-$v_{4}$ correlations at  $\sqrt{s_{NN}}$  =  200 GeV is obtained with inclusion of the system's nonlinear response to initial eccentricities accompanied by the viscous effect with $\eta/s$ $>$  0.08. 
Theoretical calculations using different initial conditions, equations of state and viscous coefficients need to be further explored to extract $\eta/s$ of the medium created at RHIC.
\end{abstract}


\end{frontmatter}

%\maketitle

\section{INTRODUCTION}
High-energy heavy-ion collisions at the Relativistic Heavy Ion Collider (RHIC) and the Large Hadron Collider (LHC) are believed to have created a QCD medium at extremely high  energy densities. The properties of this medium are under persistent investigations. 
An extensively studied subject is ``flow", the collective anisotropic  expansion of the medium. Flow originates
from the initial spatial anisotropy  (and/or fluctuations in the position) of the colliding nucleons, and develops  due to the strong interaction among particles produced in the collision~\cite{v2_1,v2_2,v2_3,v2_4,starv2_1,v2_review}.
The anisotropy in the momentum space is quantified by the Fourier coefficients of the azimuthal emission distribution of produced particles with respect to the $n^{th}$ order event plane~\cite{method,method2}:
\be
\frac{dN}{d \varphi} \propto %\frac{1}{2 \pi} \left( 
1 + \sum_{n = 1} ^\infty 2 v_n \cos \left[n(\varphi-\Psi_{n})\right] %\right)
,
\ee
where $\varphi$ is the azimuthal particle emission angle and $\Psi_{n}$ is $n^{th}$ order event plane angle. These $v_{n}$ are the single-particle Fourier coefficients that can be derived without the determination of the event plane which is discussed later on.
The flow harmonics, $v_{n}$, quantify the $n^{\rm th}$ order anisotropy of particles of interest, and its magnitude imprints the initial anisotropy~\cite{v2_1}, the expansion dynamics~\cite{hydro2,hydro3} and the equation of state of the medium~\cite{v2_2,hydro_Eos}.  
Various efforts have been made to understand the measured flow harmonics and extract the transport properties of the medium, 
but different models need a different value of $\eta/s$ (viscosity over entropy density) to describe the same experimental measurements~\cite{nexus,music}.
For example, the NeXSPheRIO model with $\eta/s$ = 0 (described in Ref.~\cite{nexus}) explains all $v_{n}$ in all centralities whereas the MUSIC model (described in Ref.~\cite{music}) indicates that a viscous medium with  $\eta/s$ $\sim$ 0.08 is needed to explain the $v_{n}$ data in Au+Au collisions at $\sqrt{s_{NN}}$  = 200 GeV. This suggests that besides the choice for $\eta/s$, other transport properties, equation of state, and the initial state also affects these calculations and using only $v_{n}$ data it is difficult to have control on these parameter.
Therefore, more inputs from experimental observables are warranted to further constrain theoretical models.

To probe the initial conditions, it is important to measure the distributions of $v_{n}$ and event-by-event correlations among $v_{n}$ values in an event sample ~\cite{v2_fluc1,v2_fluc2}
as the event-by-event correlations between different orders of flow harmonics are theorized to be sensitive to the transport  properties of the medium~\cite{v2_corr1,scmn_3}.
Recently, the four-particle symmetric cumulants~\cite{scmn_1,scmn_2,scmn_4,scmn_5,scmn_cms} have been proposed to unravel the initial-stage phenomena and the later-stage medium properties. 
These four-particle symmetric cumulants $SC(m,n)$  are defined as~\cite{scmn_1}
\begin{eqnarray}
SC(m,n) &\equiv&
\left<\left<\cos(m\varphi_1\!+\!n\varphi_2\!-\!m\varphi_3-\!n\varphi_4)\right>\right>_c\nonumber\\ 
&=& \left<\left<\cos(m\varphi_1\!+\!n\varphi_2\!-\!m\varphi_3-\!n\varphi_4)\right>\right>
-\left<\left<\cos[m(\varphi_1\!-\!\varphi_2)]\right>\right>\left<\left<\cos[n(\varphi_1\!-\!\varphi_2)]\right>\right>\nonumber\\
&=&\left<v_{m}^2v_{n}^2\right>-\left<v_{m}^2\right>\left<v_{n}^2\right>.%\nonumber\\
%&=&0\,.
\label{scmn}
\end{eqnarray}
Here subscript $c$ is used to indicate the cumulant and $\left< \right>$ denotes average over all events weighted with the number of quadruplet (doublet) combinations. The $\left<\left< \right>\right>$ denotes the average over all distinct particle quadruplets (doublets) in an event and over all events weighted with the number of quadruplet (doublet) combinations. Positive (negative) values of $SC(m,n)$ suggest the (anti-)correlation between  $v_{n}^{2}$ and $v_{m}^{2}$; in other words, $v_{n}^{2}$ being larger than $\langle v_{n}^{2}\rangle$ in an event enhances (suppresses) the probability of $v_{m}^{2}$  being larger than $\langle v_{m}^{2}\rangle$ in that same event.  The $SC(m,n)$ observables focus on the correlations between different orders of flow harmonics, and facilitate the quantitative comparison between experimental data and model calculations. A normalized symmetric cumulant will facilitate a quantitative comparison between different collision energies or between data and model calculations as the magnitude of symmetric cumulant depends on magnitude of flow harmonics. The normalized symmetric cumulant, $NSC(m,n)$, is defined as
\begin{equation}
NSC(m,n)= \frac{\langle v_{n}^{2}v_{m}^{2} \rangle - \langle
  v_{n}^{2}\rangle  \langle v_{m}^{2} \rangle}{\langle
  v_{n}^{2}\rangle  \langle v_{m}^{2} \rangle}. 
\label{scmn_norm}
%\label{eq:snnorm}
\end{equation}

The ALICE collaboration has recently measured $SC(2,3)$ and $SC(2,4)$ in Pb+Pb collisions at $\sqrt{s_{NN}}$  = 2.76 TeV, and found that the centrality dependence of $SC(2,4)$ cannot
be captured by hydrodynamics
model calculations with a constant $\eta/s$~\cite{scmn_2}.
In this Letter, we present the first measurements of symmetric cumulants in Au+Au collisions at $\sqrt{s_{NN}}$  = 39 and 200 GeV. Due to limited statistics, results from Au+Au collisions at $\sqrt{s_{NN}}$  = 62.4 GeV are not presented in this letter. Section 2 discusses experiment details and the analysis method, and Section 3 describes the results for $SC(2,3)$ and $SC(2,4)$ as a functions of centrality and compares them with available model predictions. The summary is in Section 4.

\section{EXPERIMENT AND ANALYSIS}
The data used in this measurement were collected from Au+Au collisions by the STAR~\cite{star} in the years 2010 (39 GeV) and 2011 (200 GeV).
We analyzed $1.1 \times 10^{8}$  and $4 \times 10^{8}$ minimum-bias Au+Au events at $\sqrt{s_{NN}}$  = 39 and 200 GeV, respectively.
The collision centrality determination was validated by comparing Monte Carlo Glauber calculations to the charged-hadron multiplicity measured with the time projection chamber (TPC) within a pseudorapidity window of $|\eta|$ $<$  0.5. The detailed procedures to obtain the simulated multiplicity are similar to that described in Ref.~\cite{glauber}.
Events were required to have collision vertex positions (in the radial direction) within 2 cm of the beam axis to reduce contributions from beam-pipe (at a radius of 4 cm) interactions, and within a limited distance from the center of the detector along the beam direction ($\pm$40 cm for the 39 GeV data set and $\pm$30 cm for the 200 GeV data set). Charged particles used in this analysis were reconstructed by the STAR TPC with $|\eta|$ $<$ 1.0. The distance of closest approach (DCA) of a track to the primary vertex was required to be less than 3 cm. We also required the number of fit points (nhits) used to reconstruct a track to be greater than 15 and the ratio of the number of fit points to maximum possible hits (nhits/hitmax) to be greater than 0.52. In addition we applied a transverse momentum cut (0.2 $<$ $p_{T}$ $<$ 2 GeV/$c$) to the charged tracks to minimize nonflow effects e.g. low and high $p_{T}$ cuts used to minimize resonance and jet contribution, respectively. These default cut settings were later varied for a systematic analysis.\\
The two- and four-particle correlations in Eq.~(\ref{scmn}) can be evaluated in terms of flow vectors~\cite{Q-cu}. The flow vector (or Q vector) for $n^{\rm th}$ harmonic is defined as  $Q_{n,p}\equiv\sum_{k=1}^M w_k^p e^{in\varphi_k}$, where $M$ is the multiplicity of an event. The weights ($w_k = w_{\mathrm{\rm eff}} w_{\varphi} $) were  applied to correct for the $p_T$-dependent efficiency ($w_{\rm eff} =\frac{1}{\textrm{\rm eff}(p_T)} $) and for imperfections in the detector acceptance ($w_{\varphi}$). 
The event-by-event $v_{n}^{2}$ and $ v_{n}^{2}v_{m}^{2}$ in Eq.~(\ref{scmn}) were calculated with the
multi-particle $Q$-cumulant method~\cite{scmn_1,Q-cu}:
\begin{equation}
 v_{n}^{2} = \left<\cos[n(\varphi_1\!-\!\varphi_2)]\right> = \frac{\Num{2}{n,-n}}{\Den{2}{n,-n}}
\label{eq:2p}
\end{equation}
and
\begin{eqnarray}
v_{n}^{2}v_{m}^{2} &=&\left<\cos(m\varphi_1\!+\!n\varphi_2\!-\!m\varphi_3-\!n\varphi_4)\right> \nonumber\\
&=&\frac{\Num{4}{n,m,-n,-m}}{\Den{4}{n,m,-n,-m}} , 
\label{eq:4p}
\end{eqnarray}
where
\begin{eqnarray}
\Num{2}{n,-n}&=&Q_{n,1} Q_{-n,1}-Q_{0,2}\,,\\
\Den{2}{n_1,n_2}&=&\Num{2}{0,0}=Q_{0,1}^2-Q_{0,2},
\label{eq:2pCorrelation}
\end{eqnarray}
\begin{eqnarray}
\Num{4}{n,m,-n,-m}&=&Q_{n,1} Q_{m,1} Q_{-n,1} Q_{-m,1}-Q_{n+m,2} Q_{-n,1} Q_{-m,1}
-Q_{m,1} Q_{0,2} Q_{-m,1}\nonumber\\
&&{}-Q_{n,1} Q_{m-n,2} Q_{-m,1}+2 Q_{m,3} Q_{-m,1}-Q_{m,1}
Q_{-n,1} Q_{n-m,2}\nonumber\\
&&{}+Q_{m-n,2} Q_{n-m,2}
-Q_{n,1} Q_{-n,1} Q_{0,2}+Q_{0,2} Q_{0,2}\nonumber\\
&&{}+2 Q_{-n,1} Q_{n,3}
-Q_{n,1} Q_{m,1} Q_{-n-m,2}+Q_{n+m,2}
Q_{-n-m,2}\nonumber\\
&&{}+2 Q_{m,1} Q_{-m,3}+2 Q_{n,1} Q_{-n,3}-6 Q_{0,4}\,, \\
\Den{4}{n,m,-n,-m}&=&\Num{4}{0,0,0,0}\nonumber\\
&=&Q_{0, 1}^4 - 6 Q_{0, 1}^2 Q_{0, 2} + 3 Q_{0, 2}^2 + 8 Q_{0, 1} Q_{0, 3} - 6 Q_{0, 4}\,
\label{eq:4pCorrelation} 
\end{eqnarray}
and
\begin{equation}
Q_{-n,p} = Q_{n,p}^*\,.
\end{equation}
%Here, $Q_{n,p}$ is the flow vector for the $n^{\rm th}$ harmonic, $Q_{n,p}\equiv\sum_{k=1}^M w_k^p e^{in\varphi_k}$, and $M$ is the multiplicity of an event. 
%The weights ($w_k = w_{\mathrm{\rm eff}} w_{\varphi} $) were  applied to correct for the $p_T$-dependent efficiency ($w_{\rm eff} =\frac{1}{\textrm{\rm eff}(p_T)} $) and for imperfections in the detector acceptance ($w_{\varphi}$). 
The weights of $M(M-1)$ and $M(M - 1)(M - 2)(M - 3)$ were used to average the 2-particle and 4-particle correlations over events (second average in Eq.~(\ref{scmn})).\\
The values of $v_{n}^{2}$ in the denominator of Eq.~(\ref{scmn_norm}) are obtained with the 2-particle correlations with a pseudorapidity gap of $|\Delta\eta| > 1.0$ between the two particles to suppress few-particle nonflow correlations~\cite{scmn_4}.  The expression of $v_{n}^{2}$ in terms of flow vector with a pseudorapidity gap can be written as
\begin{equation}
v_{n}^{2} =\frac{Q_{n,1}^{A}.Q_{n,1}^{B*}}{M_{A}.M_{B}}.
\label{etasub_2p}
\end{equation}
Here $Q_{n,1}^{A}$ and $Q_{n,1}^{B}$ are the flow vectors from sub-events $A$ and $B$, with $M_{A}$ and $M_{B}$ the corresponding multiplicities. Although, the eta-gap suppresses few-particle nonflow contributions (mainly due to short range correlation), nonflow due to long range correlations might affect the magnitude of measured normalized SC(m,n).\\
 Reference~\cite{scmn_6} shows that if $NSC(m, n)$ is measured in a wide centrality range, where the multiplicity significantly fluctuates, the measurements of the symmetric cumulants will be biased by such fluctuations. This is known as the centrality-bin-width (CBW) effect. Accordingly, in this analysis, the symmetric cumulants were measured in small multiplicity windows (bin size equal to one) and then combined into 10\% centrality bins to reduce statistical uncertainties. Note that we have checked, using the AMPT model, the magnitude of $NSC(m,n)$  remains unchanged if we use impact parameter bins instead of multiplicity.  \\
The main systematic uncertainties came from 1) event and track selection cuts, and 2) corrections for the non-uniform azimuthal acceptance and efficiency. 
Two methods were adopted to correct for the azimuthal dependence of the tracking efficiency: $\varphi$-weighting and re-centering~\cite{method}. In the $\varphi$-weighting method, each particle is weighted by the inverse of the corresponding efficiency, $w_{\varphi}$, determined from the particle azimuthal distribution (averaged over many events). In the re-centering method, the event-averaged Q-vector is subtracted from the Q-vector of each event and then the same equations as described above are used. In this analysis, the re-centering correction was applied as a function of multiplicity.  Both the $\varphi$-weighting and the re-centering  methods were applied separately for each run period of data taking and centrality interval. The difference between the two correction methods was included in the systematic uncertainty. To estimate systematic uncertainty due to variation of tracks and event selection cuts, we have varied DCA, nhits, nhits/hitmax and Vz values from default cut value. 
The total systematic uncertainty was obtained by adding uncertainties from different sources in quadrature. Maximum contribution ($\sim$8-12$\%$) to the systematic uncertainty comes from the correction for non-uniform azimuthal acceptance and efficiency. 

\section{RESULTS}
%%%%%%%%%%%%%% Fig. 1 %%%%%%%%%%%%%%%%%%%%%%%%%%%%%
\begin{figure}[h]
\begin{center}
\begin{overpic}[scale=0.4]{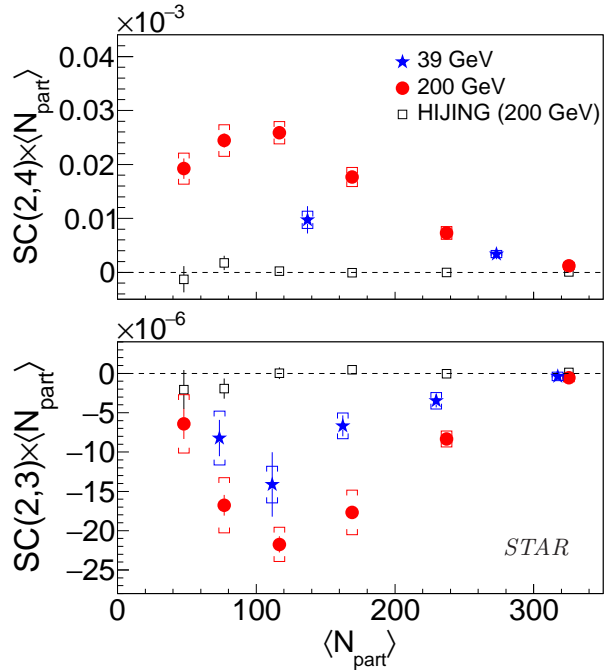}
\put (70,20) {\small \it STAR}
\end{overpic}
\caption{(Color online) $SC(2,4)\times \langle N_{\rm part} \rangle$ and  $SC(2,3) \times \langle N_{\rm part} \rangle$ as functions of $\langle N_{\rm part} \rangle$  in Au+Au collisions at $\sqrt{s_{NN}}$ = 39 and 200 GeV. HIJING model results are shown only for 200 GeV. Verticals lines are statistical uncertainties and systematic errors are shown with cap symbols.}
\label{fig1_scmn}
\end{center}
\end{figure}

%%%%%%%%%% End of Fig.1 %%%%%%%%%%%%%%%%%%%%%%%%%%%%%
Figure~\ref{fig1_scmn} presents the symmetric cumulants  $SC(2,3)$ and  $SC(2,4)$ multiplied by the average number of participating nucleons, $\langle N_{\textrm{part}} \rangle$,  as functions of centrality (represented by $\langle N_{\rm part} \rangle$~\cite{glauber}) at midrapidity ($|\eta| < 1.0$) for charged hadrons in Au+Au collisions at $\sqrt{s_{NN}}$ = 39 and 200 GeV. 
Systematic errors are shown with cap symbols. Due to limited statistics at 39 GeV, the $SC(2,4)$ is measured in two wide centrality bins, 0-20$\%$ and 20-40$\%$.
Positive values of $SC(2,4)$ are observed for all centrality intervals at both collision energies, suggestive of a correlation between $v_2$ and $v_4$.  On the other hand, the negative values of $SC(2,3)$ reveal the anti-correlation between $v_2$ and $v_3$. The magnitude of the $SC(2,3)\times\langle N_{\textrm{part}} \rangle$ and $SC(2,4)\times\langle N_{\textrm{part}} \rangle$  increases from central to mid-peripheral events and then again decreases for very peripheral events. This is also true for $SC(2,3)$ and  $SC(2,4)$ (not shown).
An inherent feature of the symmetric cumulant is the suppression of nonflow effects thanks to the use of the four-particle cumulant, where nonflow refers to azimuthal correlations not related to the reaction plane orientation, arising from resonances, jets, quantum statistics, and so on.
Figure~\ref{fig1_scmn} also shows HIJING model calculations of $SC(2,3)$ and  $SC(2,4)$  in Au+Au collisions at $\sqrt{s_{NN}}$ = 200 GeV~\cite{hijing1,hijing2}. 
Comparison to the HIJING model, which includes only nonflow physics, suggests that nonflow effects cannot explain the data of non-zero symmetric cumulants.\\
\bef
\begin{center}
\begin{overpic}[scale=0.4]{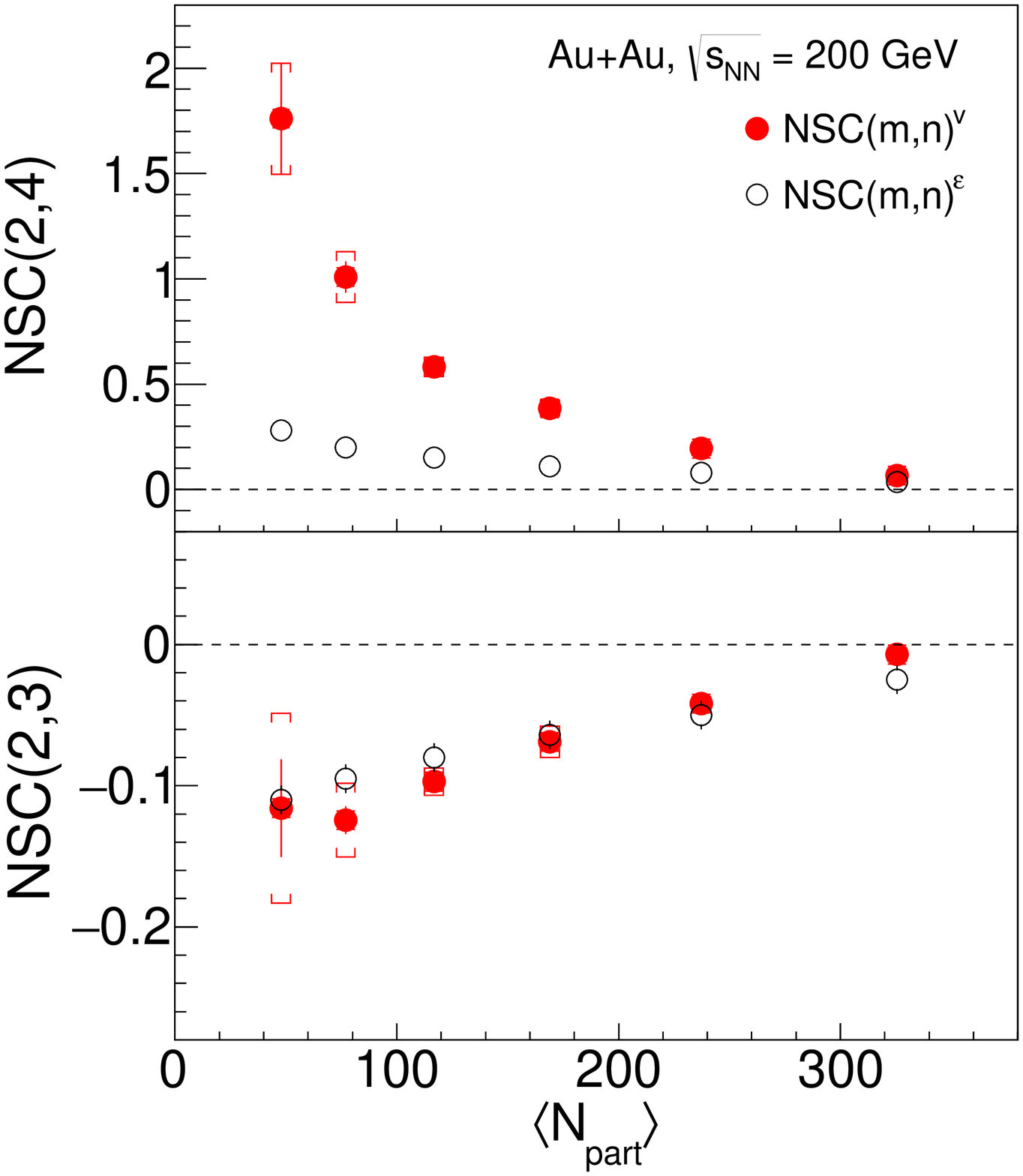}
\put (70,20) {\small \it STAR}
\end{overpic}
\caption{(Color online) $NSC(2,4)$ 
and  $NSC(2,3)$ as functions of $\langle N_{\rm part} \rangle $ in Au+Au collisions at $\sqrt{s_{NN}}$ = 200 GeV.  The normalized symmetric cumulants of spatial eccentricities, $NSC(m,n)^{\varepsilon}$,  using the Glauber model, are also shown.  $NSC(m,n)^{v}$ represent  normalized symmetric cumulants of flow coefficients using equation~\ref{scmn_norm}. 
}
\label{fig_ee_scmn}
\end{center}
\eef
Anisotropic flow is generated by the initial geometric anisotropy coupled with a collective expansion of the produced medium. There is an intense interest in understanding the origin of the initial-stage fluctuations and how these fluctuations manifest themselves in correlations between measured particles.
The normalized symmetric cumulants evaluated in the coordinate space, $NSC(m,n)^{\varepsilon} = (\left< \varepsilon_{n}^2 \varepsilon_{m}^2 \right> - \left< \varepsilon_{n}^2 \right> \left< \varepsilon_{m}^2 \right>)/(\left<\varepsilon_{n}^2\right>\left<\varepsilon_{m}^2\right>)$,  in Au+Au collisions at $\sqrt{s_{NN}}$ = 200 GeV (using the Monte Carlo Glauber model~\cite{scmn_6}) are shown in Fig.~\ref{fig_ee_scmn} and compared with $NSC(m,n)^{v}$ measured in the momentum space~\cite{method}. If only eccentricity drives $v_{n}$, then we expect $NSC(m,n)^{v}$ = $NSC(m,n)^{\varepsilon}$. Figure~\ref{fig_ee_scmn} demonstrates that the initial-stage anti-correlation between the $2^{\rm nd}$ ($\varepsilon_{2}$) and $3^{\rm rd}$ ($\varepsilon_{3}$) order eccentricity is mainly responsible for the observed anti-correlation between $v_{2}$ and $v_{3}$. However, the correlation between $\varepsilon_{2}$ and $\varepsilon_{4}$  under-predicts the observed correlation between $v_{2}$ and $v_{4}$. The difference between $NSC(2,4)^{v}$  and $NSC(2,4)^{\varepsilon}$ increases from central to peripheral collisions.  The anisotropic flow $v_4$ has a contribution not only from the linear response of the system to $\varepsilon_{4}$, but also has a contribution proportional to $\varepsilon_{2}^{2}$. 
 Therefore, the increased difference between $NSC(2,4)^{v}$  and $NSC(2,4)^{\varepsilon}$ from central to peripheral collisions is presumably because $\varepsilon_{2}$ has an increased contribution in $v_{4}$ in more-peripheral collisions. This is consistent with the observation reported by the ATLAS~\cite{v2_corr_atlas}  and ALICE~\cite{scmn_2} experiments in Pb+Pb collisions at $\sqrt{s_{NN}}$ = 2.76 TeV.  The relative contribution of $\varepsilon_{2}$ in $v_{4}$, as compared to that of $\varepsilon_{4}$  was suggested to depend on the viscous properties of the medium~\cite{non_linear}. Therefore, $NSC(2,4)$  provides a probe into the medium properties.

We present the collision energy dependence of the normalized symmetric cumulants  $NSC(2,4)$  and  $NSC(2,3)$  in Fig.~\ref{fig_nscmn}, as functions of $\langle N_{\rm part} \rangle$.
 The magnitude of $NSC(2,4)$ is systematically higher at the lower energy (39 GeV) compared with 200 GeV and 2.76 TeV, though the observed difference is not statistically significant (a $\sim$2$\sigma$ effect). This difference could be related to the change in the initial conditions and/or in the transport properties of the medium with collision energy~\cite{non_linear,scmn_6}. Future high-statistics measurements of $NSC(2,4)$ at low energies in the phase II of the Beam Energy Scan program (BES-II) at RHIC will further our understanding of the temperature dependence of $\eta/s$. The lower panel of  Fig.~\ref{fig_nscmn}  shows  a comparison of $NSC(2,3)$  as a function of centrality between 39 GeV and 200 GeV. The results of $NSC(2,3)$ at 39 and 200 GeV are consistent with each other. 
In Fig.~\ref{fig_nscmn}, the ALICE measurements for 2.76 TeV Pb+Pb~\cite{scmn_2} are also shown for comparison. Since the ALICE results did not take into account the CBW effect, we have also illustrated the STAR results without the CBW correction (labeled as ``Wide Mult. Bins"  in Fig.~\ref{fig_nscmn}) at 200 GeV to make a fair comparison.  There is a slight difference in $NSC(2,4)$ between results with and without the CBW correction; however, this effect is larger in $NSC(2,3)$. 
The uncorrected values of $NSC(2,3)$ or $NSC(2,4)$ at 200 GeV and 2.76 TeV are very close to each other for all centrality intervals.  

\bef
\begin{center}
\begin{overpic}[scale=0.4]{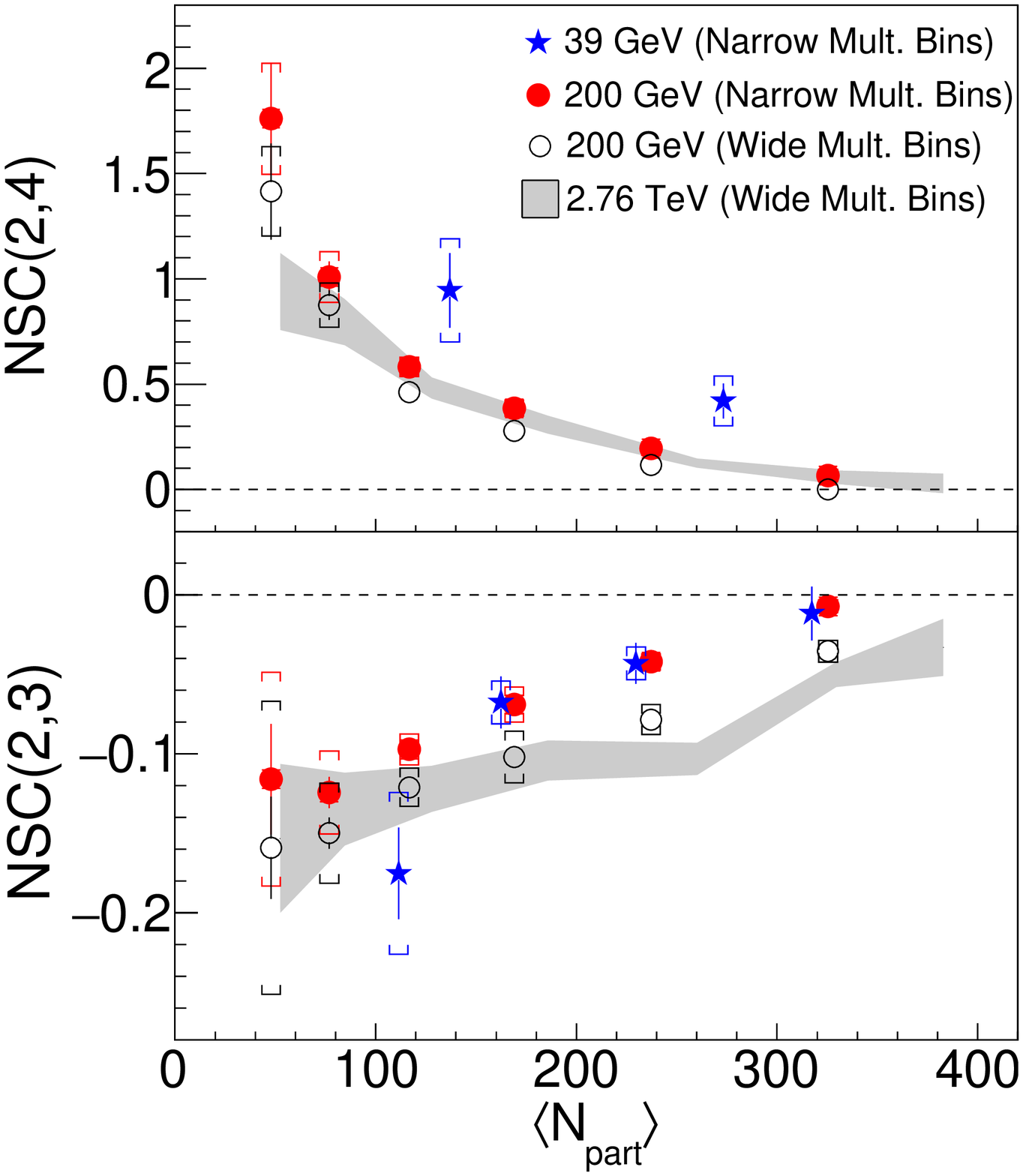}
\put (70,20) {\small \it STAR}
\end{overpic}
\caption{(Color online)  $NSC(2,4)$ 
and  $NSC(2,3)$ as functions of average $\langle N_{\rm part} \rangle$ in Au+Au collisions at $\sqrt{s_{NN}}$ = 39 and 200 GeV.  ALICE results for 2.76 TeV Pb+Pb~\cite{scmn_2} are also shown for comparison.  Vertical lines are statistical uncertainties.
Systematic errors are shown with cap symbols. The CBW corrected results are labeled by ``Narrow Mult. Bins" and CBW uncorrected results are labeled by ``Wide Mult. Bins".
}
\label{fig_nscmn}
\end{center}
\eef

We compare in Fig.~\ref{fig4_scmn} our measurements with available model predictions~\cite{scmn_5,scmn_6} for $NSC(2,3)$ and $NSC(2,4)$ in Au+Au collisions at 200 GeV.
The AMPT calculations~\cite{scmn_5} are shown for a partonic medium with  $\eta/s$ = $0.18$ (i.e., 3 mb parton-parton interaction cross-section). In the AMPT model, $\eta/s$ in partonic matter is estimated with the assumption that the partonic matter only consists of  massless u and d quarks~\cite{ampt_eta_by_s}. AMPT model calculations with $\eta/s$ = $0.18$ are in agreement with the $v_{n}$ magnitudes in peripheral collisions, but it over-predicts  the $v_{n}$ data for the most-central collisions~\cite{scmn_5}. The $NSC(2,3)$ and $NSC(2,4)$ from data are reasonably well described by the AMPT model, however an increasing deviation between data and AMPT model calculation is observed for $NSC(2,4)$ at peripheral collisions. Predictions from an ideal hydrodynamics model (NexSPheRIO)~\cite{scmn_6} are also shown in Fig.~\ref{fig4_scmn}. The ideal hydrodynamics model with NexSPheRIO initial conditions
is able to explain the anti-correlation between $v_{2}$ and $v_{3}$  within theoretical uncertainties, but under-predicts the correlation between $v_{2}$ and $v_{4}$. The NexSPheRIO model describes the magnitudes of all flow harmonics ($v_{2}$, $v_{3}$ and $v_{4}$) up to $p_{T}$ $\sim$ 2.0 GeV/c within 10$\%$, measured in all centrality intervals at the top RHIC energy~\cite{scmn_6}. The failure of the ideal hydrodynamics model for the $v_{2}$-$v_{4}$ correlation supports the idea that the symmetric cumulants provide additional constraints to theoretical models.
Like the ideal hydrodynamics model, a viscous hydrodynamics model (with MCKLN initial condition and $\eta/s=0.08$) roughly explains the $NSC(2,3)$ data and under-predicts $NSC(2,4)$  for peripheral collisions. However, the prediction from the viscous hydrodynamics model for $NSC(2,4)$ is closer to data than the ideal hydrodynamics model. 
Note that, all presented models (hydro and transport) under-predict $NSC(2,4)$. 
This may be improved by revisiting model ingredients such as the average initial state, initial state fluctuations, energy deposition  ``smearing", equation of state, the appearance of other forms of transport, etc. Hence, a sound conclusion requires further investigation along that line. 
\bef
\begin{center}
\begin{overpic}[scale=0.4]{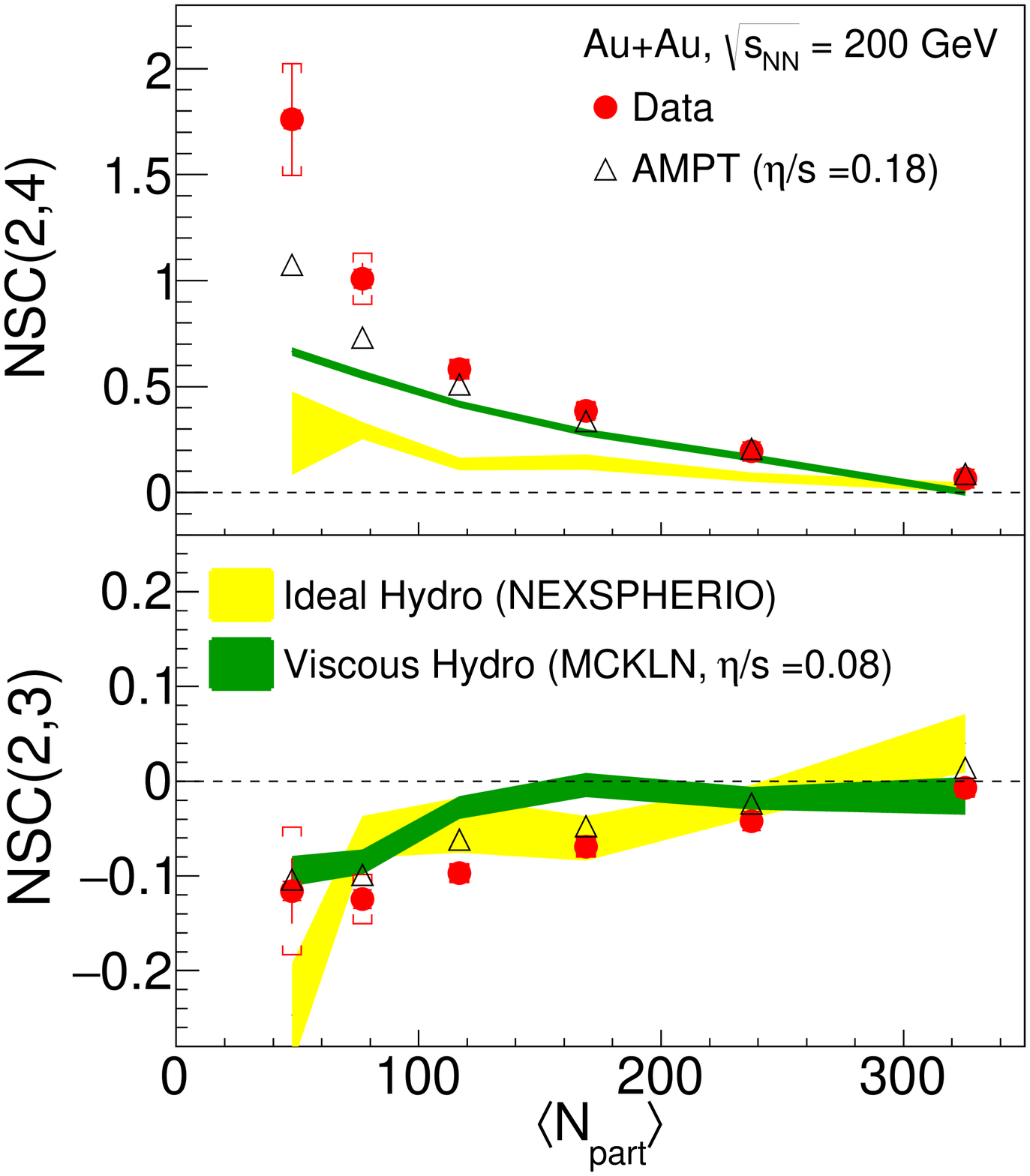}
\put (70,20) {\small \it STAR}
\end{overpic}
\caption{(Color online) $NSC(2,4)$ 
and  $NSC(2,3)$ as functions of average $\langle N_{\rm part} \rangle$  in Au+Au collisions at $\sqrt{s_{NN}}$ = 200 GeV, with hydrodynamics~\cite{scmn_6} and AMPT model~\cite{scmn_5} predictions for comparison. All results are CBW corrected.
}
\label{fig4_scmn}
\end{center}
\eef

\section{CONCLUSIONS}
We have presented the first measurements of the charge-inclusive four-particle symmetric cumulants as functions of centrality at midrapidity in Au+Au collisions at $\sqrt{s_{NN}}$  = 39 and 200 GeV.
The  new data provide additional constraints on the initial conditions and the transport properties in theoretical models.
Anti-correlation has been observed between event-by-event fluctuations of $v_{2}$ and  $v_{3}$, while the event-by-event fluctuations of $v_{2}$ and  $v_{4}$ are found to be correlated.  
The initial-stage anti-correlation between $\varepsilon_2$ and $\varepsilon_3$ appears to describe the observed anti-correlation between $v_{2}$ and $v_{3}$, which seems to support the idea of the linearity between $\varepsilon_n$ and $v_{n}$~\cite{method}. 
However, the initial-stage correlation alone is not sufficient to describe the measured correlation between $v_{2}$ and  $v_{4}$: the nonlinear hydrodynamic response of the medium has to be included to reproduce the data. The $v_{2}$-$v_{4}$ correlation seems to be different between 200 GeV and 39 GeV, which could be attributed to the corresponding difference in the initial conditions and/or the transport properties of the medium.
We have compared the STAR measurements with a number of available theoretical model calculations. All the models explain the symmetric cumulant between $v_2$ and $v_3$; however, none of them are able to describe the $v_2$-$v_4$ correlation for all centralities (though the hydrodynamics model ($\eta/s$ = 0) could successfully reproduce the individual flow harmonics). 
A viscous hydrodynamics model with $\eta/s$ = 0.08 outperforms the ideal hydrodynamics model in explaining the data, but still seems to under-predict $NSC(2,4)$, whereas the AMPT calculations with $\eta/s$ = 0.18 are even closer to the measurement. A detailed comparison of the presented data to models with different equation of states, initial conditions, and transport coefficients is needed to determine these coefficients quantitatively.

%\noindent{\bf Acknowledgments}\\
\section*{Acknowledgement}
We thank the RHIC Operations Group and RCF at BNL, the NERSC Center at LBNL, and the Open Science Grid consortium for providing resources and support. This work was supported in part by the Office of Nuclear Physics within the U.S. DOE Office of Science, the U.S. National Science Foundation, the Ministry of Education and Science of the Russian Federation, National Natural Science Foundation of China, Chinese Academy of Science, the Ministry of Science and Technology of China and the Chinese Ministry of Education, the National Research Foundation of Korea, GA and MSMT of the Czech Republic, Department of Atomic Energy and Department of Science and Technology of the Government of India; the National Science Centre of Poland, National Research Foundation, the Ministry of Science, Education and Sports of the Republic of Croatia, RosAtom of Russia and German Bundesministerium fur Bildung, Wissenschaft, Forschung and Technologie (BMBF) and the Helmholtz Association.

\section*{References}
%\bibliography{mybibfile}
%\normalsize

\end{document}

%% file: author.tex
\author{
J.~Adam$^{9}$,
L.~Adamczyk$^{1}$,
J.~R.~Adams$^{30}$,
J.~K.~Adkins$^{20}$,
G.~Agakishiev$^{18}$,
M.~M.~Aggarwal$^{32}$,
Z.~Ahammed$^{55}$,
N.~N.~Ajitanand$^{43}$,
I.~Alekseev$^{16,27}$,
D.~M.~Anderson$^{45}$
% ...
\\
(STAR Collaboration)
}
\address{$^{1}$AGH University of Science and Technology, FPACS, Cracow 30-059, Poland}
% ...
\address{$^{9}$Creighton University, Omaha, Nebraska 68178}
% ...
\address{$^{16}$Alikhanov Institute for Theoretical and Experimental Physics, Moscow 117218, Russia}
% ...
\address{$^{18}$Joint Institute for Nuclear Research, Dubna, 141 980, Russia}
% ...
\address{$^{20}$University of Kentucky, Lexington, Kentucky 40506-0055}
% ...
\address{$^{27}$National Research Nuclear University MEPhI, Moscow 115409, Russia}
% ...
\address{$^{30}$Ohio State University, Columbus, Ohio 43210}
% ...
\address{$^{32}$Panjab University, Chandigarh 160014, India}
% ...
\address{$^{43}$State University of New York, Stony Brook, New York 11794}
% ...
\address{$^{45}$Texas A\&M University, College Station, Texas 77843}
% ...
\address{$^{55}$Variable Energy Cyclotron Centre, Kolkata 700064, India}
% ...
